\documentclass[aps,prb,twocolumn,showpacs,groupeaddress]{revtex4}

\usepackage{indentfirst}
\usepackage{graphicx}
\usepackage{subfigure}
\usepackage{amsmath}

\begin{document}

\title{FORC Analysis of homogeneous nucleation in the two-dimensional kinetic Ising model}

\author{D.T. Robb}
\email{robb@csit.fsu.edu}
\affiliation{School of Computational Science, Florida State University, Tallahassee, FL 32306}
\affiliation{ERC Center for Computational Sciences, Mississippi State University, Mississippi State, MS 39762}
\author{M.A. Novotny}
\email{novotny@erc.msstate.edu}
\affiliation{Department of Physics and Astronomy and ERC Center for Computational Sciences, Mississippi State University, Mississippi State, MS 39762}
\author{P.A. Rikvold}
\email{rikvold@csit.fsu.edu}
\affiliation{School of Computational Science, Center for Materials Research and Technology, and Department of Physics, Florida State University, Tallahassee, FL 32306}
\affiliation{National High Magnetic Field Laboratory, Tallahassee, FL 32310}

\date{\today}

\begin{abstract}

The first-order reversal curve (FORC) method is applied
to the two-dimensional kinetic Ising model. 
For the system size and magnetic field chosen, the system reverses by
the homogeneous nucleation and growth of \emph{many} droplets. This makes the
dynamics of reversal nearly deterministic, in contrast to
the strongly disordered systems previously studied by the FORC
method. Consequently, the FORC diagrams appear different from
those obtained in previous studies.
The Kolmogorov-Johnson-Mehl-Avrami (KJMA) theory of phase 
transformation by nucleation and growth is applied to the system. Reasonable 
agreement with the Monte Carlo simulations is found, and the FORC method 
suggests how the KJMA theory could be extended. 

\end{abstract}


\maketitle

\section{Introduction}

The study of hysteresis loops can produce significant dynamical 
information, but it reduces the often complex dynamics of 
magnetization reversal to only a few quantities, usually the 
hysteresis loop area
and the coercive field. The first-order reversal curve (FORC) method
was recently developed \cite{kn:pike99a} to extract more information from 
experiments on magnetic systems. It has produced interesting results, 
mostly in systems with strong disorder in the physics of magnetism 
\cite{kn:bercoff02, kn:katzgraber02, kn:pike03} and 
in the geosciences. \cite{kn:pike01a, kn:muxworthy02}

\section{FORC analysis and Monte Carlo simulation}

\begin{figure*}[ht]
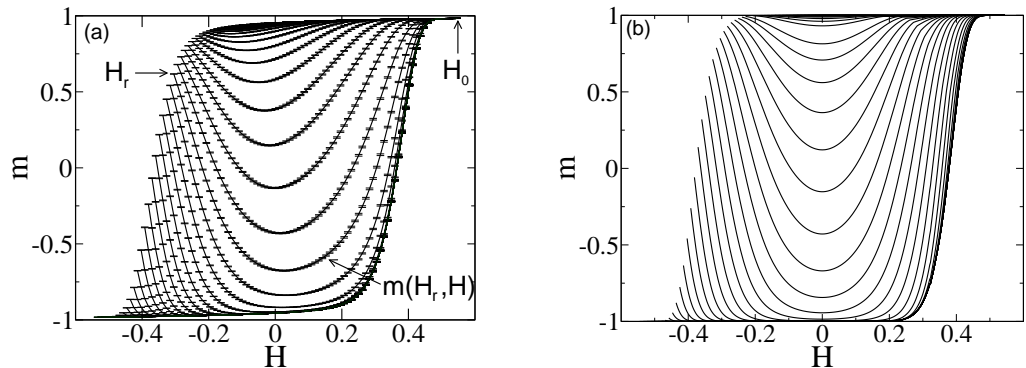

\mbox{
 \subfigure
{\includegraphics[height=1.9in]{forcfam.512.t1000.label.eps}} \qquad \quad
 \subfigure
 {\includegraphics[height=1.9in]{forcfam.kjma.t1000.eps}}} 
\caption
{\label{forcfams} Family of FORCs for $L = 512$ kinetic Ising model at temperature $T = 0.8 T_c$ in a linearly varying field with sweep rate $\Omega = 2.18 
\times 10^{-3} \; \mathrm{MCSS}^{-1}$. (a) Data from MC simulation. 
(b) Calculation from application of KJMA theory.} 
\end{figure*}

The FORC technique involves saturating the magnetization $M$ in a positive 
field $H_0$, decreasing the field to a series of progressively more negative
return fields $H_r$, and then in each case increasing the field back to 
$H_0$ while recording the magnetization $m = M/M_s$, where $M_s$ is the
saturation magnetization. This
results in a set of curves $m(H_r,H)$, where $H$ represents the field
as it is increased from $H_r$ back to $H_0$, as shown in 
Fig.~\ref{forcfams}(a).

We use FORC analysis to better understand the process of hysteresis
in the two-dimensional Ising model, which is defined by the Hamiltonian
\begin{equation}
{\cal H } = - J \sum_{\langle i, j \rangle} S_i S_j - \mu H(t) \sum_i S_i ;
\label{Ising1}
\end{equation}
where the lattice spins $S_i = \pm 1$, the exchange constant $J$ is in units of energy, $\mu$ is the 
magnetic moment per site, and $H(t)$ is the applied time-dependent field. The
simulation is run on an $L \times L$ square lattice with periodic boundary
conditions. The 
first sum is taken over all nearest-neighbor pairs of spins, and the second
sum over all individual spins. Hereafter we use units such that 
$J = \mu = k_B = 1$. Our simulations are run at a temperature of $T=0.8 T_c$,
and we use the Glauber dynamic with random, single-site 
updates for the Monte Carlo (MC) simulation. \cite{kn:landau00} The
field decreases linearly at the sweep rate $- \Omega $ from
$H_0$ to $H_r$, and then increases linearly at the sweep rate $+ \Omega $ back
to $H_0$.

The free-energy barrier for the metastable state in the Ising system leads either to a co-existence (CE), single-droplet (SD), multi-droplet (MD), or 
strong-field (SF) regime for magnetization reversal. \cite{kn:rikvold94a}
Occurring at intermediate values of system size, field and 
temperature, the MD regime is arguably the most accessible to experiments. 
The SF regime
requires fields that may be too large to generate experimentally, while
the SD regime requires small systems or very low temperatures, and thus
very long metastable lifetimes. 
To ensure that the reversal is in the MD regime, the saturation field is chosen
as $H_0 = 0.545$, and the system size as $L = 512$. We observed directly that at
$L = 512$ more than 100 droplets were involved in the reversal, and
our FORC results did not change appreciably for larger L or larger $H_0$.
We calculated the characteristic reversal time 
in a constant field $H = -H_0$, defined as the time to decrease from 
$m=1.0$ to $m=-0.8$, to be roughly 100 MCSS (MC steps per spin). 
We therefore chose $\Omega = 2.18 \times 10^{-3} \; \mathrm{MCSS}^{-1}$, so 
that the time for the field to decrease from $H = 0$ to 
$H = - H_0$ was long enough to produce full reversal.  In 
Fig.~\ref{forcfams}(a) we present the resulting family of FORCs, where each
FORC is an average over 20 MC realizations.

It is useful to generate from this family of FORCs the FORC distribution,
\begin{equation}
\rho(H_r,H) = - \frac{1}{2}
\frac{\partial^2 m(H_r,H)}{\partial H_r \partial H}.
\label{eqFORC}
\end{equation}
The motivation for this definition is that when the FORCs are taken from a 
Preisach system, the FORC
distribution reproduces the Preisach distribution. \cite{kn:pike99a}
In order to calculate the FORC distribution from our MC data, 
as in Ref. \onlinecite{kn:pike99a} we fit the family of curves $m(H_r,H)$ around
each point by a second-order polynomial $g(H_r,H) = a H_r H + b H_r^2 + c H^2
+ d H_r + e H + f$, using a linear least-squares method.  The FORC 
distribution is most
clearly displayed in a contour plot of its level curves, called a `FORC
diagram,' which we show in Fig.~\ref{forcdiagrams}(a).

There are several interesting features of the FORC results. 
First, a sizable part of the FORC diagram
is negative. Since the FORC distribution measures how the slope $ \partial m / 
\partial H$ varies
from one FORC to the next, which can be quite different for different dynamics,
this is a reasonable result. It
is only in the case of the Preisach model that the FORC distribution must by definition be positive. Second, the positions
of the minima of the FORCs, naively expected to be at $H = 0$, 
gradually shift from a small negative field to a small positive
field. We will provide insight into these features in section \ref{s:kjma}.

\section{Application of the KJMA theory}  \label{s:kjma}

We briefly sketch the idea of the Kolmogorov-Johnson-Mehl-Avrami (KJMA) theory.
See Refs.~\onlinecite{kn:rikvold94a} and \onlinecite{kn:sides99} for more
complete descriptions. 
In the present case of MD reversal, the system parameters 
are related as $a \ll r_c \ll r_0 \ll L$, where these parameters represent 
the lattice spacing, the critical droplet radius, the average distance 
before droplets meet during growth, and the
system size, respectively. The critical droplet radius is the radius below 
which droplets of overturned spins tend strongly to shrink, and above which 
they tend strongly to grow.

For the MD regime, magnetization reversal can be approximated by the KJMA theory
of droplet nucleation and growth, in which critical droplets are assumed to 
nucleate and grow with field- and temperature-dependent rates. First, the 
extended volume $\Phi (t)$ of the expanding droplets, uncorrected for droplet
overlap, is calculated. Then the expression $\phi (t) = \exp [-\Phi (t)]$
is applied to find the volume $\phi (t)$ of overturned spins, corrected for 
droplet overlap. 
(See Ref.~\onlinecite{kn:sides99} and references cited therein.)

To apply the KJMA theory, we included the linearly varying field in 
the KJMA expressions, found straightforward equations for the extended
volume in the region of negative fields, and then numerically integrated these
equations. As in Sides et al., \cite{kn:sides99} we approximated the behavior 
of the system in positive fields as a combination
of a shrinking negative metastable region, whose volume is estimated as 
the time reversal of the calculated growth in negative
fields, and nucleation of a positive stable phase within this region, 
estimated from the calculated nucleation in
negative fields.  Because the calculated growth in negative fields extends 
only to $H = H_r$, this approximation can continue only up to
$H = |H_r|$. However, the system at $H = |H_r|$,
where positive droplets are growing, can be approximated by a similar
(but oppositely magnetized) state of the system, obtained from the
full hysteresis loop. We found the point in the calculation
of the full hysteresis loop where the magnetization is equal in magnitude to,
but opposite in sign from, the magnetization at $H = |H_r|$. Beginning with 
the results of the KJMA integration at this point, we continued the 
integration from $H = |H_r|$ to $H = H_0$. 

\begin{figure*}[ht]
\mbox{
 \subfigure
{\includegraphics[angle=0,height=1.8in]{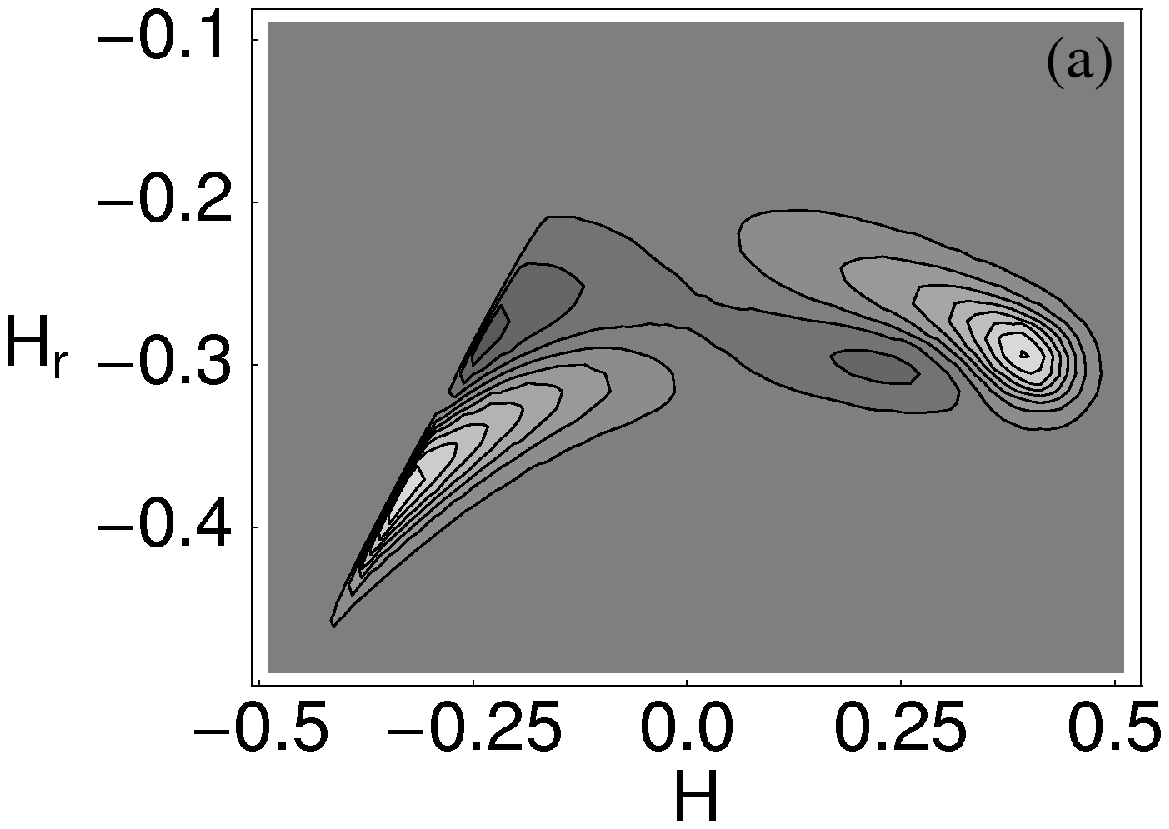}} \qquad
 \subfigure
 {\includegraphics[angle=0,height=1.8in]{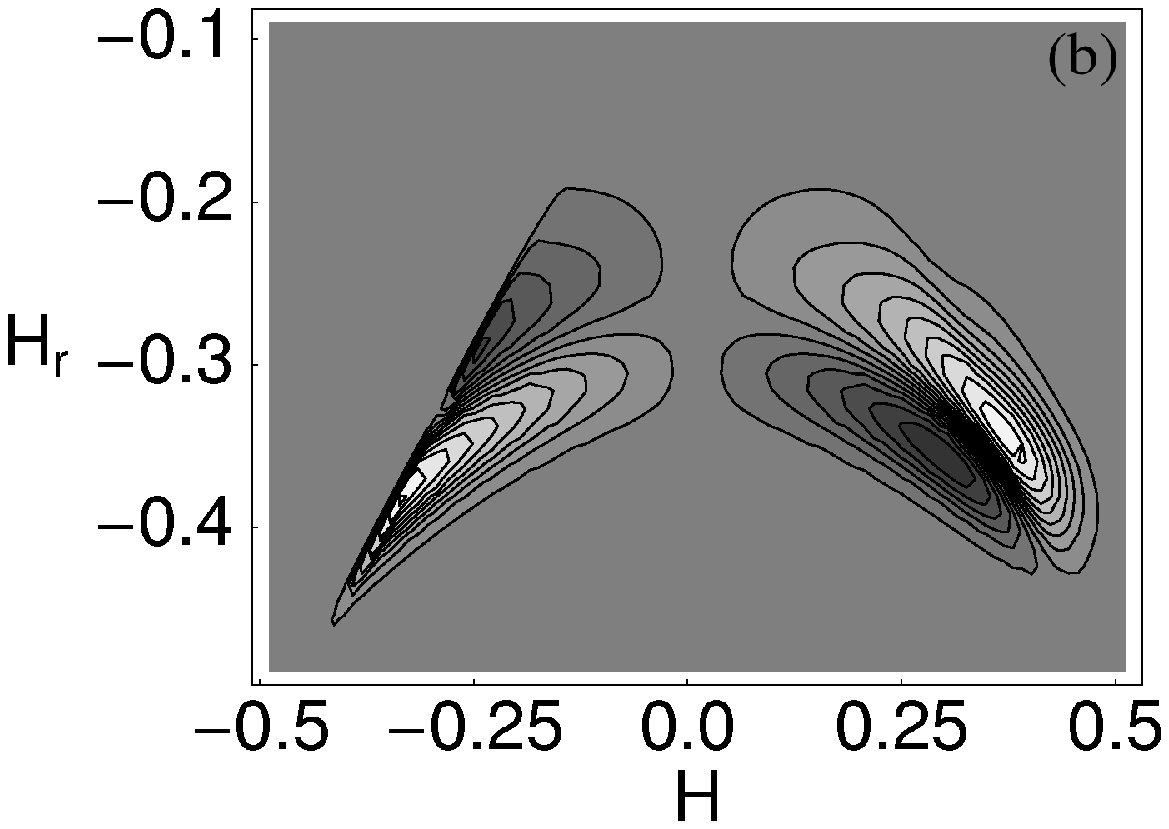}}}
\caption
{ \label{forcdiagrams} FORC diagrams for the system described in 
Fig.~\ref{forcfams}. The magnitudes in the contour plot range from 
$\rho = -400$ (black) to $\rho = 0$ (large gray area) to 
$\rho = +400$ (white). 
(a) Calculated from MC simulation data. 
(b) Calculated with KJMA theory.} 
\end{figure*}

In comparing the calculated results to the MC simulation data, we note first
that the approximation of
treating the final portion of each FORC loop using results from the KJMA
simulation in negative fields works well, deviating only in small kinks 
(almost invisible on the scale used here) which develop
in the first FORCs in Fig.~\ref{forcfams}(b). This suggests that the
behavior of the remaining positive metastable phase in 
negative field values is similar to the behavior of the remaining negative
metastable phase in the last stage of the FORCs. The first FORCs are 
least similar, probably because they consist of several well-defined
shrinking droplets, rather than an irregularly spaced and shaped volume. It
is also interesting that the magnetization decreases from saturation
much sooner in the MC family of FORCs, due to the presence of subcritical 
droplets which are neglected in the KJMA treatment. This causes the values
of the FORC distribution to be larger for the KJMA case, since the slopes of
the FORCs must be higher.

We produced animations of the MC simulation in order to better understand the dynamics.
In Fig.~\ref{forcfams}(a), the FORCs beginning in the range 
$-0.2 > H_r > -0.25$ have their minima at $H < 0$ 
because droplets that are just above the critical radius become subcritical
and shrink as the field increases back toward $H = 0$. The
FORCs beginning in the range $-0.25 > H_r > -0.32$ have their minima 
at values $H>0$ for a similar reason. These minima occur at $m<0$, where the
magnetization can be viewed as consisting of positive droplets in a negative 
background. In the subsequent small positive fields, the growth of the larger
positive droplets is accompanied by the shrinking of the smaller, subcritical 
positive droplets. The entire shift of the minima is reflected in the presence 
of values $\rho < 0$ of the FORC distribution in the region 
$(-0.30<H_r<-0.25,-0.05<H<0.05)$ in Fig.~\ref{forcdiagrams}(a).  

In areas of the FORCs in which a full reversal
occurred, there is a loss of memory in the MC simulation, resulting 
in a value $\rho \approx 0$ for these areas in Fig.~\ref{forcdiagrams}(a). 
We also studied the families of
FORCs and FORC diagrams for the same system with $\Omega = 2.18
\times 10^{-4} \; \mathrm{MCSS}^{-1}$ (not shown). 
We found the form of the MC and KJMA results were quite similar to the
$\Omega = 2.18 \times 10^{-3} \;  \mathrm{MCSS}^{-1}$ case, except that the 
coercive field was about half as large.

\section{Conclusion} 

In conclusion, we have applied FORC analysis to the
two-dimensional kinetic Ising model, finding reasonable agreement
between the predictions of KJMA theory and the results
of MC simulation. The FORC analysis
offers insight into the discrepancies between the KJMA and MC
results, which we have identified as due to the effects of subcritical 
droplets and loss of memory.

In future work, we hope to extend the KJMA theory to include these
effects and to make predictions with the KJMA theory for longer
time scales, where MC simulation is not practical. Although the simulation
results for $\Omega = 2.18 \times 10^{-3} \; \mathrm{MCSS}^{-1}$ and 
$\Omega = 2.18 \times 10^{-4} \; \mathrm{MCSS}^{-1}$ were very similar, the long
time scales of physical systems may produce different and physically relevant
insights into the dynamics of the homogeneous reversal process.
 
We thank H.G. Katzgraber and C.R. Pike for useful comments.
This work was supported by NSF Grant No. DMR-0120310, by the ERC Center for
Computational Sciences at Mississippi State University, and by the School of
Computational Science at Florida State University.

%
%

\end{document}